\documentclass{icrc2009}

\usepackage{graphicx}   
\usepackage[small]{caption}    
\usepackage{subfig} 
\usepackage{fixltx2e}
\usepackage{url}

\newcommand{\shorttitle}[1]%
{\markboth{Proceedings of the 31\MakeLowercase{$^{st}$} ICRC, {\L}\'{o}d\'{z} 2009}{#1} }
\newcommand{\etal}{\MakeLowercase{\textit{et al. }}} 

\begin{document}
\title{Search for Individual UHECR Sources in the Future Data}

\author{\IEEEauthorblockN{Gwenael Giacinti\IEEEauthorrefmark{1} and
			  Dmitri V. Semikoz\IEEEauthorrefmark{1}}
                            \\
\IEEEauthorblockA{\IEEEauthorrefmark{1}AstroParticle and Cosmology (APC), 10 rue Alice Domon et L\'eonie Duquet, 75205 Paris Cedex 13, France}}

\shorttitle{Giacinti \etal Search for UHECR sources}
\maketitle

\begin{abstract}
We propose a new way to detect individual bright Ultra-High Energy Cosmic Ray (UHECR) sources above background if the Galactic Magnetic Field (GMF) gives the main contribution to UHECR deflections~\cite{SearchForSources}. This method can be directly applied to maps given by experiments. It consists in starting from at least two high energy events above 6\,$\times$\,10$^{19}$\,eV, and looking at lower energy tails. We test the efficiency of the method and investigate its dependence on different parameters. In case of detection, the source position and the local GMF deflection power are reconstructed. Both reconstructions are strongly affected by the turbulent GMF. With the parameters adopted in this study, for 68\% of reconstructed sources, the angular position is less than one degree from the real one. For typical turbulent field strengths of 4\,$\mu$G at the Earth position and 1.5\,kpc extension in the halo, one can reconstruct the deflection power with 25\% precision in 68\% of cases.
  \end{abstract}


\section{Introduction}

The HiRes experiment first observed the cutoff in the spectrum~\cite{spectrum_HiRes}. The UHECR sources are still unknown. Some possible candidates can be studied from the point of view of acceleration mechanisms -see for example Refs.~\cite{Hillas:1985is,Ptitsyna:2008zs,Neronov:2007mh}.\\
Their detection is difficult due to the UHECR deflections in the extragalactic and Galactic magnetic fields. One can either wait for enough data at the highest energies to find a class of UHECR sources, or search for the first individual brightest sources. In this study, we investigate this second possibility.\\
We propose here a method to find Ultra-High Energy (UHE) proton or light nuclei sources on top of background and reconstruct their positions. We show that one has to start from at least two events with energies above $10^{19.8}$\,eV\,$\sim 6 \times 10^{19}$\,eV before looking at lower energy tails, in order to avoid confusion by the background. We investigate the performance of this method depending on the different relevant parameters.\\
We assume as in~\cite{Dolag:2004kp} that UHECR deflections due to the extragalactic magnetic fields are negligible compared to those due to the GMF. For an example of non negligible extragalactic contributions, see~\cite{Sigl:2004yk}. The GMF is divided in two components: regular and turbulent. For the regular one, we take the Prouza and Smida model~\cite{PS,GMF}. It consists of a thick disk, and of toroidal and dipolar fields. Following the latest knowledge on the GMF~\cite{Waelkens:2008gp,Sun:2007mx}, we modify some parameters and take an exponentially decaying profile along the Galactocentric radius. The field strength close to the Sun is $B_{\rm reg}=1-3\,\mu$G (currently admitted strength: $B_{\rm reg} \simeq 2\,\mu$G). The model for the turbulent component is described in~\cite{SearchForSources}. The root mean square (RMS) value of the turbulent field decays as~$1/r$ along the Galactocentric radius~$r$, and exponentially along the direction orthogonal to the Galactic plane. The plane thickness is set here to 1.5\,kpc. The turbulent field strength close to the Sun is $B_{\rm turb}=2-8\,\mu$G (currently admitted strength: $B_{\rm turb} \simeq 4\,\mu$G).\\

 \begin{figure}[!t]
  \centering
  \includegraphics[width=0.5\textwidth]{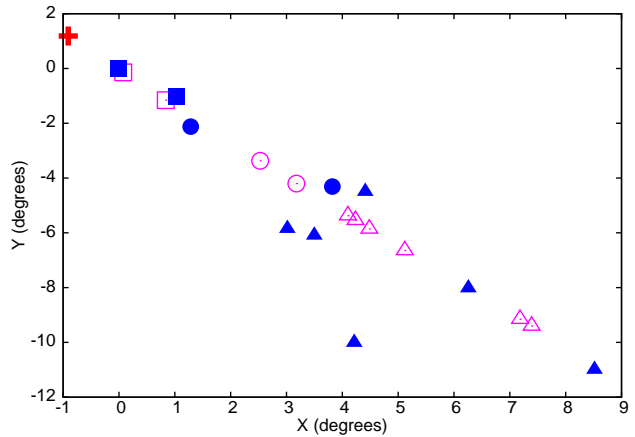}
  \caption{Projections of the arrival directions of cosmic rays (CR) emitted by a source (cross) in the plane tangent to the celestial sphere, and centered on the highest energy event (filled square in (0,0)) emitted by the source. Deflections X and Y measured on two orthogonal axes, in degrees. Open symbols for source CR deflected by the regular GMF only. Filled symbols for the same CR when the turbulent component is also taken into account. Shapes correspond to CR energies (triangles: between $10^{19.0}$ and $10^{19.3}$\,eV, circles: between $10^{19.3}$ and $10^{19.6}$\,eV, and squares: above $10^{19.6}$\,eV).}
  \label{TurbulentComponent}
 \end{figure}

For source events, we assume power law acceleration spectra, with a power law index equal to 2.2 and a maximum energy set to 10$^{21}$\,eV. In most cases, we take a proton source and set its distance to the Earth to $50-100$\,Mpc. Protons are propagated to the observer by using the results of Ref.~\cite{Kachelriess:2007bp} for energy losses, which creates a ``bump'' in the source spectrum~\cite{bump}. Then, the generated cosmic rays (CR) are deflected in the regular GMF (see e.g.~\cite{Harari2}), and in the turbulent GMF. For more details on turbulent GMF deflections, see Refs.~\cite{TurbulentComponent1,SearchForSources}. The effects of both components are shown in Fig.~\ref{TurbulentComponent}: The regular component deflects the CR along a curve and the turbulent one spreads them ``randomly'' around it in a sector-shaped region. When there are many events with the same energies, they are not spread uniformly, as discussed in References~\cite{Harari2,TurbulentComponent1,Golup:2009zg}.\\
Results depend on many parameters of the model and on the considered location on the sky. However, in practice, there are only two essential parameters: The local deflection powers -see Ref.~\cite{SearchForSources}- of the regular GMF (denoted $D$) and of the turbulent GMF.\\
Background events are simulated according to the energy spectrum measured by HiRes~\cite{spectrum_HiRes} and the exposure of Telescope Array~\cite{exposure}.

\section{Method}
\label{sectionmethod}

\begin{figure}[!t]
  \centering
  \includegraphics[width=0.5\textwidth]{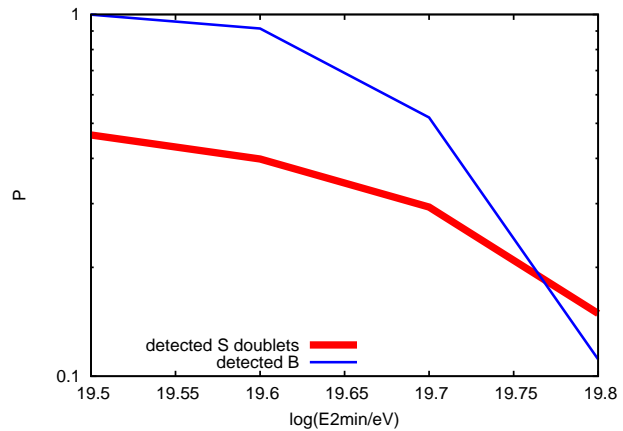}
  \caption{Dependence on $E_{\rm 2min}$, the minimum energy for the second highest energy event. Thick solid line for the probability to detect the source. Thin solid line for the probability to detect some background. Results presented for a source of luminosity above 10$^{19.8}$\,eV equal to 2.6\% of the total luminosity in UHECR, $E_{\rm th}=10^{19.3}$\,eV, 5000 events above 10$^{19}$\,eV in the whole sky, $B_{\rm reg}=2\,\mu$G and $B_{\rm turb}=4\,\mu$G.}
  \label{ShE}
 \end{figure}

The following work is done on sky maps which only display events with energies bigger than a given threshold, $E_{\rm th}$. The idea is to look for tails of lower energy events around at least two nearby events with energies $E>10^{19.8}$\,eV.\\
We start by taking a circle around an event with $E_{1}>10^{19.8}$\,eV. Its radius $R$ can be optimized and the value we take depends, among others, on $E_{\rm th}$ and $D$. We will call $R$ an ``internal'' parameter of the method. The tail of lower energy events is searched by assuming that all events from the source are located in a sub-region of the circle, that has a sector shape with a central axis given by the second highest energy event. The energy $E_{2} \leq E_{1}$ of the second highest energy event is above a given threshold $E_{2} \geq E_{\rm 2min}$, and its distance to the highest energy event must be compatible with an emission from the same source. This distance should be lower than $\beta D/E_{2} - D/E_{1}$, where $\beta$ is another internal parameter, and $D$ is the value initially assumed for the local deflection power. Fig.~\ref{ShE} shows the probability to detect the source and the probability to be confused by the background, for different values of $E_{\rm 2min}$. One can see that one has to require $E_{\rm 2min}=10^{19.8}$\,eV, in order not to detect some background more frequently than the source. That is why we search for at least two nearby events at the highest energies, $E>10^{19.8}$\,eV.\\

Then, as a discriminator of source events, we use a correlation coefficient -see Ref.~\cite{Golup:2009zg}. One takes the energies $E$ and the coordinates $X'$ of the events located in the sector, where $X'$ is the axis containing the first event and the center of mass of the other considered events. Each event is tested and if removing it increases the correlation coefficient $C(1/E,X')$, it is then definitely removed. Otherwise, it is definitely kept. If $C(1/E,X')$ is finally larger than a given value, $C_{\min}$, there is detection: Detection of the source in case of an initial source doublet or confusion by the background in case of an initial background doublet.\\
Ref.~\cite{SearchForSources} discusses the optimization of the internal parameters.\\
In case of source detection, we finally reconstruct its position along the $X'$ axis, by taking the remaining events and fitting $1/E$ versus $X'$ with a straight line. This also gives an estimate of the local regular GMF deflection power, $D$.

\section{Study of the parameter space}
\label{sectiondependance}

We study below how the efficiency of the method presented above changes with values of physical or experimental parameters.\\
There are two requirements for source detection. First, the probability to detect the source doublets has to be large enough, in order not to remove too much source signal. Second, the probability to be confused by some background has to be low enough compared to the probability to detect the source signal.\\
We define here a ``cluster'' as a group of at least three nearby events with very high energies, $E>10^{19.8}$\,eV. The probability to have a cluster of background events is very low in case of reasonable statistics and no -or low- anisotropy. For example it is $\sim 1$\%, with 5000 events in the whole sky above 10$^{19}$\,eV and no anisotropy. On the contrary, the probability to have a background doublet of very high energy events is, in most cases, not negligible compared to the probability to have a source doublet. That is why we separate the two cases -doublet or cluster- below. In case of a doublet, we apply the method depicted in the previous section, whereas in case of a cluster, we do not need it since the source is already detected in $\sim 99$\% of cases. Applying the method reduces the background more than the signal from the source, but its side-effect is to reduce the signal from the source in any case. That is why it is not worth applying it when the probability to be confused by some background is already very low. However, when statistics increase, or in case of a large anisotropy above $6 \times 10^{19}$\,eV, background clusters become more frequent and the method would then also be interesting for clusters of 3 events.\\

 \begin{figure}[!t]
  \centering
  \includegraphics[width=0.5\textwidth]{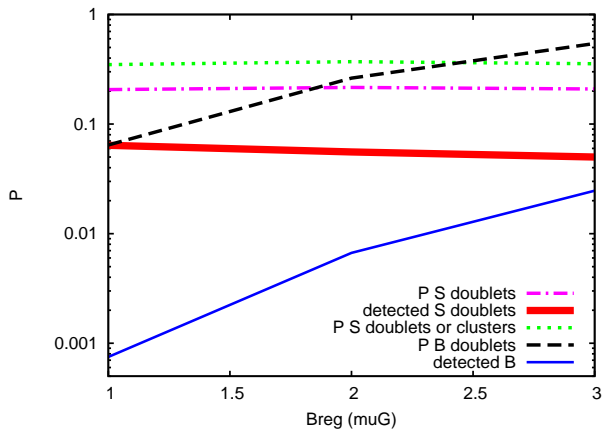}
  \caption{Dependence on the regular GMF strength, $B_{\rm reg}$, with a constant ratio $B_{\rm reg}/B_{\rm turb}$ set to 0.5. $E_{\rm th}=10^{19.6}$\,eV. Source luminosity and number of events on the sky as in Fig.~\ref{ShE}. Dotted line for the probability to have a doublet or a cluster from the source. Dashed-dotted and dashed lines for the probabilities to have a source or background doublet, respectively. Same key for solid lines as in Fig.~\ref{ShE}.}
  \label{RegField}
 \end{figure}

 \begin{figure}[!t]
  \centering
  \includegraphics[width=0.5\textwidth]{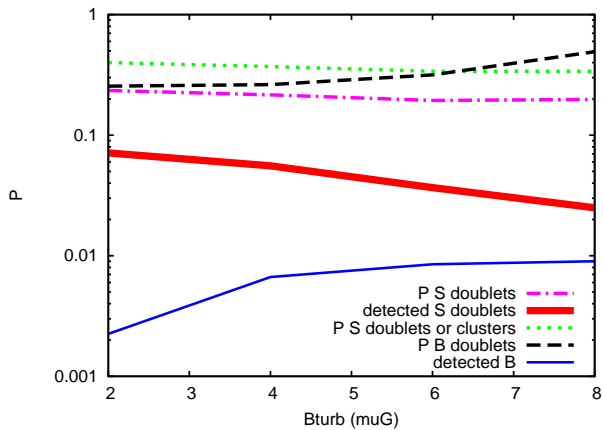}
  \caption{Dependence on the turbulent GMF strength, $B_{\rm turb}$, with a constant regular field set to $B_{\rm reg}=2\,\mu$G. Same values for the other parameters and same line types as in Fig.~\ref{RegField}.}
  \label{TurbField}
 \end{figure}

The figures~\ref{RegField} and~\ref{TurbField} show the dependence of results on the regular and turbulent GMF strengths. Fig.~\ref{RegField} is computed for different strengths of $B_{\rm reg}$ between 1 and 3\,$\mu$G, and with a constant ratio between regular and turbulent field strengths. For Fig.~\ref{TurbField}, $B_{\rm reg}=2\,\mu$G and the turbulent field strength $B_{\rm turb}$ varies between 2 and 8\,$\mu$G. The internal parameters of the method have been optimized for each field strength. The dotted line is the probability that a proton source of luminosity above 10$^{19.8}$\,eV equal to 2.6\% of the total luminosity in UHECR emits a doublet or a cluster of events with energies $E>10^{19.8}$\,eV. The dashed-dotted and the dashed lines are respectively the probabilities to have a source or background doublet. The thick and the thin solid lines represent the probabilities to respectively detect the source or the background. One can notice that even for the largest field strengths in these ranges, the ability of the method to detect the source is still quite acceptable. The difference between the dashed and the thin solid lines shows the efficiency to remove the background. The difference between the dashed-dotted and the thick solid lines is smaller: The source signal is less removed than the background. The difference between dotted and dashed-dotted lines corresponds, in practice, to cases when the source is already detected thanks to clusters of events.\\
These results are computed for an isotropic sky, even at the highest energies. We checked that for reasonable values of anisotropies (if all events with energies above 10$^{19.8}$\,eV are located in a fraction of the sky larger than 25\%), results would be affected in a linear way. For example, in the case of a 25\% fraction, the probabilities to have a background doublet and to detect it would be both approximately multiplied by 4.\\
We have studied the dependence on source parameters. We show in Ref.~\cite{SearchForSources} results for the dependence on the source luminosity. We point out that with a regular GMF deflection power, for example close to $D \sim 5^{\circ} \times 10^{19.6}$\,eV for protons, one can still detect a light nuclei source, but only by looking for clusters if the source luminosity is sufficient. In this case, the ratio ``signal/background'' for doublets would be too small to detect such sources. We have also presented results on the dependence on experimental statistics. There is no clear and no general best energy threshold $E_{\rm th}$ to look for lower energy tails of events. When $E_{\rm th}$ is decreased, the probability to detect a source doublet through the method increases, but the probability to be confused by a background doublet increases with a larger rate. For the example considered in Ref.~\cite{SearchForSources}, $E_{\rm th}=10^{19.6}$\,eV\,$\simeq 4 \times 10^{19}$\,eV is a good compromise between sufficient source detection and sufficient background rejection. The impact of the total amount of experimental data mostly affects the way the detection should be done, depending on the source luminosity. For example, with a bright source and large statistics like 10$^{4}$ events above 10$^{19}$eV, clusters of source events appear more often and looking for them becomes a more efficient way to track the source than applying the method for doublets. However, when the amount of data still increases, the method used previously for doublets will also start to be needed for clusters, so as to reject background ones.

\section{Reconstruction of the source position and of the regular GMF deflection power}
\label{sectionreconstruction}

In this section, we discuss the ability of the method to reconstruct the source position and the local regular GMF deflection power $D$.

 \begin{figure}[!t]
  \centering
  \includegraphics[width=0.5\textwidth]{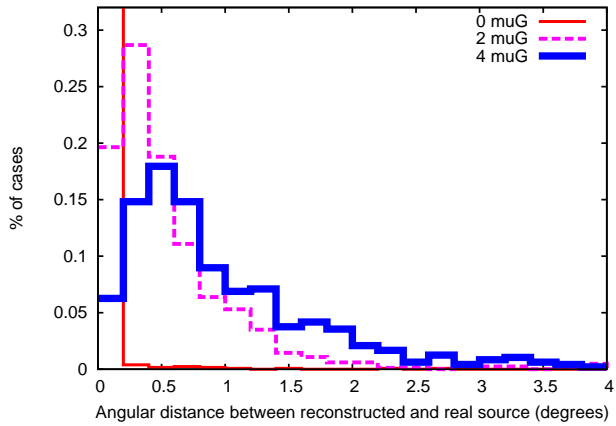}
  \caption{Distribution of distances, in degrees, between the reconstructed sources and the real source. The source is located in a region of the sky where $D \sim 5^{\circ} \times 10^{19.6}$\,eV. Thin solid line for no turbulent component ($B_{\rm turb}=0\,\mu$G), dashed line for $B_{\rm turb}=2\,\mu$G, and thick solid line for $B_{\rm turb}=4\,\mu$G.}
  \label{Distances}
 \end{figure}

 \begin{figure}[!t]
  \centering
  \includegraphics[width=0.5\textwidth]{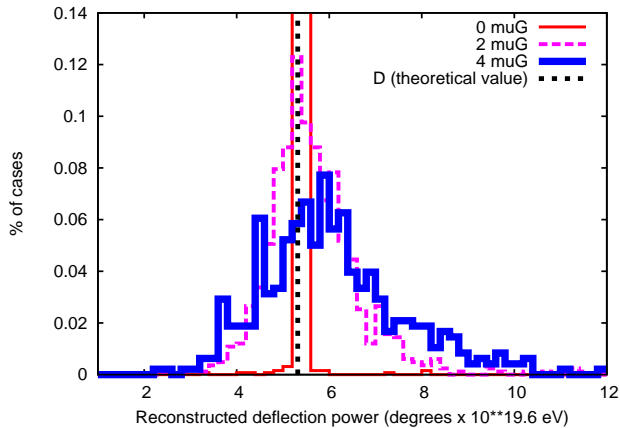}
  \caption{Distribution of reconstructed deflection powers in~$^{\circ} \times 10^{19.6}$\,eV. Same parameters and line types as in Fig.~\ref{Distances}.}
  \label{DPow}
 \end{figure}

With the assumptions considered here, and $B_{\rm turb}=4\,\mu$G, the position of a source located in a region of the sky where $D \sim 5^{\circ} \times 10^{19.6}$\,eV can be reconstructed with an accuracy of 1 degree in about 68\% of cases -see distribution with thick solid line in Fig.~\ref{Distances}. Thus, the angular resolution of this method for the source position reconstruction is of the order of the experimental angular resolution.\\
Another noticeable result is that, in the same conditions, the local deflection power $D$ of the regular GMF component can be reconstructed with up to 25\% precision in 68\% of cases, as shown in Fig.~\ref{DPow} -see thick solid line.\\
The above results are mainly spoiled by the turbulent field. For comparison, we put in the same plots the distributions with $B_{\rm turb}=0\,\mu$G (thin solid line) and $B_{\rm turb}=2\,\mu$G (dashed line). The lower the turbulent field strength, the better the precisions on both results. For these lower $B_{\rm turb}$ strengths, the distance on the sky between the reconstructed and the real sources would be mostly affected by the experimental angular resolution and not by the method.\\
The strength of the regular field (for $B_{\rm reg}=1-3\,\mu$G at the Sun position) does not change the precision on the reconstruction of $D$, but the precision on the source position reconstruction is better for low $B_{\rm reg}$ values.\\
If we simulate the experimental resolution on energies by taking values such as $\Delta E/E \sim 20$\% instead of $\Delta E/E=0$ (theoretical energies), no noticeable change can be seen in our results.

\section{Conclusion}

We have proposed and studied here an alternative way to detect individual bright UHECR sources, if the UHECR deflections due to the turbulent GMF are not too large compared to the deflections due to the regular component. We generated two sets of events: Events from a source with a $1/E^{2.2}$ injection spectrum that we propagated from the source to the Earth, and events from some background generated according to the HiRes spectrum and the Telescope Array exposure. We mixed both sets of events, tried to detect the source and checked how often we got confused by detecting some background.\\
We showed that one should look for doublets of events with $E>10^{19.8}$\,eV, when using the depicted method.\\
We studied in section~\ref{sectiondependance} the dependence on several parameters: Unknown parameters from the magnetic fields, the source, the sky anisotropy at high energies and the experimental statistics. We showed in section~\ref{sectionreconstruction} that with the parameters and assumptions adopted here, the precision on the source reconstruction would be dominated by the current experimental angular resolution in 68\% of cases. The local deflection power can be known up to 25\% in 68\% of cases.\\
In the future, this method can be applied to experimental data so as to find bright UHECR proton or light nuclei sources.


\begin{thebibliography}{99}

\bibitem{SearchForSources}
  G.~Giacinti, X.~Derkx and D.~V.~Semikoz,
  JCAP {\bf 1003}, 022 (2010)
  [arXiv:0907.1035 [astro-ph.HE]].

\bibitem{spectrum_HiRes}
  R.~Abbasi {\it et al.}  [HiRes Collaboration],
  Phys.\ Rev.\ Lett.\  {\bf 100}, 101101 (2008)
  [arXiv:astro-ph/0703099].

 \bibitem{Hillas:1985is}
   A.~M.~Hillas,
    Ann.\ Rev.\ Astron.\ Astrophys.\  {\bf 22}, 425 (1984).

\bibitem{Ptitsyna:2008zs}
  K.~Ptitsyna and S.~V.~Troitsky,
  Phys.\ Usp.\  {\bf 53}, 691 (2010)
  [arXiv:0808.0367 [astro-ph]].

\bibitem{Neronov:2007mh}
  A.~Y.~Neronov, D.~V.~Semikoz and I.~I.~Tkachev,
  New J.\ Phys.\  {\bf 11}, 065015 (2009).
  [arXiv:0712.1737 [astro-ph]].

\bibitem{Dolag:2004kp}
  K.~Dolag, D.~Grasso, V.~Springel and I.~Tkachev,
  JCAP {\bf 0501}, 009 (2005)
  [arXiv:astro-ph/0410419].

\bibitem{Sigl:2004yk}
  G.~Sigl, F.~Miniati and T.~A.~Ensslin,
  Phys.\ Rev.\  D {\bf 70}, 043007 (2004)
  [arXiv:astro-ph/0401084].

\bibitem{PS}
  M.~Prouza and R.~Smida,
  Astron.\ Astrophys.\  {\bf 410} (2003) 1
  [arXiv:astro-ph/0307165].

\bibitem{GMF}
  M.~Kachelriess, P.~D.~Serpico and M.~Teshima,
  Astropart.\ Phys.\  {\bf 26} (2006) 378
  [arXiv:astro-ph/0510444].

\bibitem{Waelkens:2008gp}
  A.~Waelkens, T.~Jaffe, M.~Reinecke, F.~S.~Kitaura and T.~A.~Ensslin,
  arXiv:0807.2262 [astro-ph].

\bibitem{Sun:2007mx}
  X.~H.~Sun, W.~Reich, A.~Waelkens and T.~Enslin,
  arXiv:0711.1572 [astro-ph].

\bibitem{Kachelriess:2007bp}
  M.~Kachelriess, E.~Parizot and D.~V.~Semikoz,
  JETP Lett.\  {\bf 88}, 553 (2009)
  [arXiv:0711.3635 [astro-ph]].

\bibitem{bump}
  V.~S.~Berezinsky and S.~I.~Grigor'eva,
  Astron.\ Astrophys.\  {\bf 199} (1988) 1.

\bibitem{Harari2}
  D.~Harari, S.~Mollerach and E.~Roulet,
  JHEP {\bf 0207} (2002) 006
  [arXiv:astro-ph/0205484].

\bibitem{TurbulentComponent1}
  D.~Harari, S.~Mollerach, E.~Roulet and F.~Sanchez,
  JHEP {\bf 0203} (2002) 045
  [arXiv:astro-ph/0202362].

\bibitem{Golup:2009zg}
  G.~Golup, D.~Harari, S.~Mollerach and E.~Roulet,
  AIP Conf.\ Proc.\  {\bf 1123}, 240 (2009)
  [arXiv:0902.1742 [astro-ph.HE]].

\bibitem{exposure}
  P.~Sommers,
  Astropart.\ Phys.\  {\bf 14} (2001) 271
  [arXiv:astro-ph/0004016].

  \end{thebibliography}
\end{document}